\documentclass[a4paper,11pt]{article}
\pdfoutput=1 

\usepackage{jinstpub} 

\usepackage{lineno}

\title{\boldmath 10 ps timing with highly irradiated 3D trench silicon pixel sensors}

\author[a,f,1]{A. Lampis,\note{Corresponding author.}}
\author[b,g]{F. Borgato,}
\author[a]{D. Brundu,}
\author[a]{A. Cardini,}
\author[a]{G.M. Cossu,}
\author[e,h]{G.-F. Dalla Betta,}
\author[a,f]{M. Garau,}
\author[a]{L. La Delfa,}
\author[a]{A. Lai,}
\author[a]{A. Loi,}
\author[d,i]{M. Obertino,}
\author[b,g]{G. Simi,}
\author[c]{S. Vecchi}

\affiliation[a]{INFN, Sezione di Cagliari, Cagliari, Italy}
\affiliation[b]{INFN, Sezione di Padova, Padova, Italy}
\affiliation[c]{INFN, Sezione di Ferrara, Ferrara, Italy}
\affiliation[d]{INFN, Sezione di Torino, Torino, Italy}
\affiliation[e]{INFN, TIFPA, Trento, Italy}
\affiliation[f]{Dipartimento di Fisica dell'Universit\`a di Cagliari, Cagliari, Italy}
\affiliation[g]{Dipartimento di Fisica dell'Universit\`a di Padova, Padova, Italy}
\affiliation[h]{Dipartimento di Ingegneria Industriale, Universit\`a di Trento, Trento, Italy}
\affiliation[i]{Dipartimento di Scienze Agrarie, Forestali ed Alimentari dell’Università di Torino, Grugliasco, Italy
}

\emailAdd{andrea.lampis@cern.ch}

\abstract{In this paper the results of a beam test characterization campaign of 3D trench silicon pixel sensors are presented. A time resolution in the order of 10 ps was measured both for non-irradiated and irradiated sensors up to a fluence of $2.5 \cdot 10^{16}\,1\,MeV\, n_{eq}\,cm^{-2}$. This feature and a detection efficiency close to $99\%$ make this sensors one of the best candidates for 4D tracking detectors in High-Energy-Physics experiments.}


\proceeding{23$^{\text{rd}}$ International Workshop on
Radiation Imaging Detectors\\
  26 – 30 June 2022\\
  Riva del Garda, Italy}

\begin{document}
\maketitle
\flushbottom

\section{Introduction}
High-Energy-Physics experiments are demanding higher and higher event rates to increase statistics that allows to access to new physics processes and to decrease statistical uncertainties. Upgraded colliders are fulfilling this demand running at increased luminosity and bringing new challenges to tracking detectors. The main problem that these detectors have to cope is a higher radiation damage up to $10^{17}\,1\,MeV\, N_{eq}\,cm^{-2}$. The increase in luminosity will also bring higher pile-up and the event reconstruction will become more difficult. The LHCb collaboration is evaluating the possibility to build a new vertex detector with time information at the hit level to recover tracking and vertexing capabilities, a 4D tracker~\cite{ftdr}. Looking to the LHCb Upgrade Phase II specifications the ideal sensor technology must have, at the same time, a spatial resolution in the order of 10~$\mu$m, a time resolution less than 50 ps and a radiation hardness from $10^{16}$ to $10^{17}\,1\,MeV\, n_{eq}\,cm^{-2}$. Recent results from the INFN funded project TimeSPOT have shown the excellent time resolution of 3D silicon sensors optimized for timing performances~\cite{jinst}\cite{jinst2}. In the next sections new beam test characterizations of these innovative 3D silicon sensors are presented, with a focus on the comparison between non-irradiated and irradiated sensors at very high fluences.
\label{sec:intro}

\section{3D sensors}
\label{sec:3dsensors}
\begin{figure}[htbp!] 
\centering    
\includegraphics[width=0.5\textwidth]{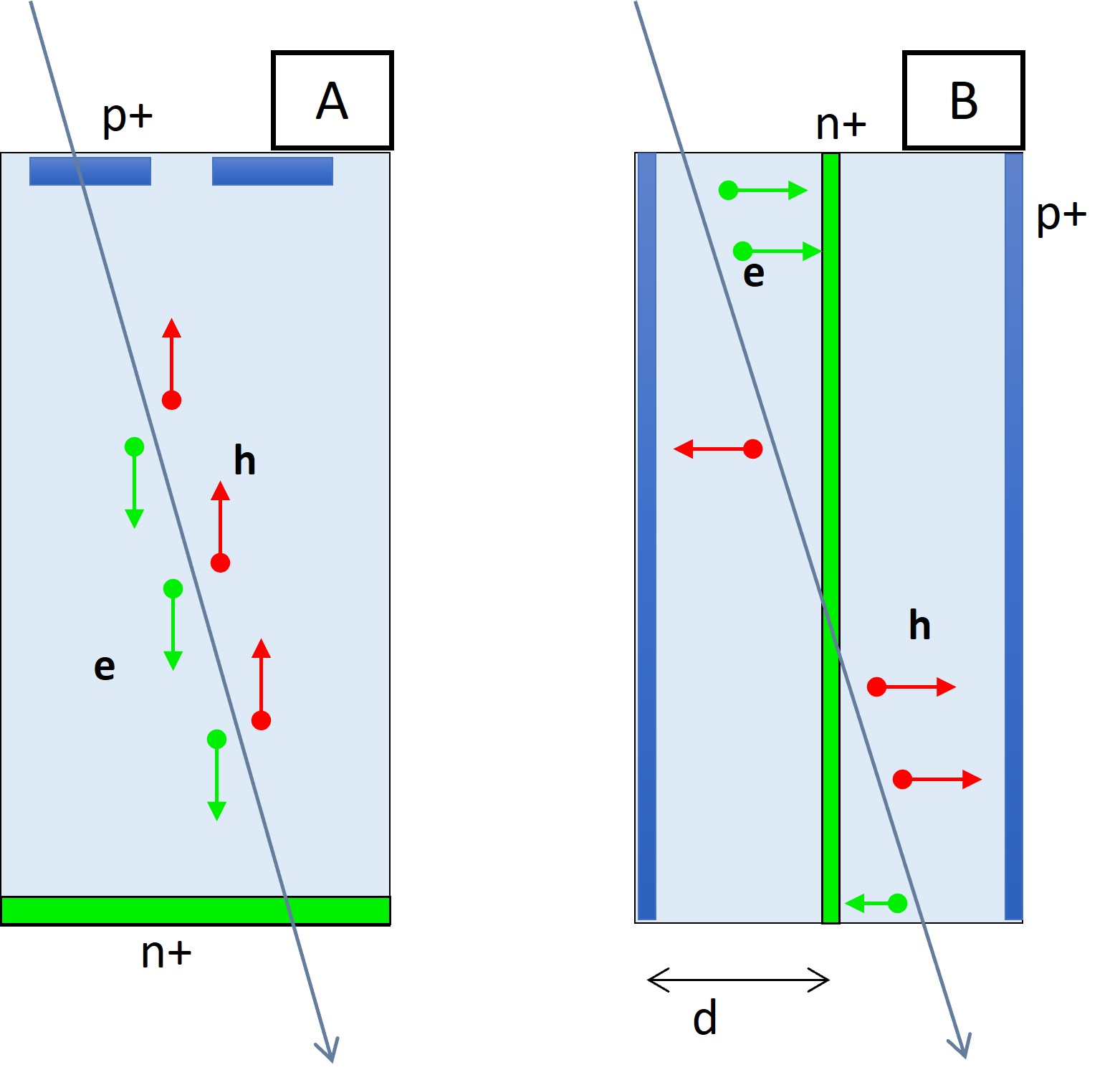}
\caption{Comparison of the charge collection process between a planar sensor (left) and a 3D sensor (right).}
\label{fig:3dvsplan}
\end{figure}

3D sensors are silicon pixel sensors with the electrodes directly built into the sensor thickness. This feature allows to develop sensors with a very short inter-electrode distance with respect to standard planar sensors, Figure~\ref{fig:3dvsplan}, without reducing the active material in which the crossing particles can create e-h pairs. This allows to exploit short inter-electrodes distances without loosing in signal amplitude.

One of the main features of these sensors are the extremely fast signals, given by the shorter path travelled by the charge carriers before being collected by the electrodes. Moreover, higher electric fields and thus charge carriers mobility saturation are obtained with a lower bias voltages. Another important advantage of sensors with short inter-electrode distance is a high radiation hardness, that for a columnar 3D silicon sensor was tested up to $1\cdot 10^{17}\,1\,MeV\, n_{eq}\,cm^{-2}$~\cite{radhard}. These two characteristics lead 3D sensors to be one of the most appealing sensors for tracking detectors for HL-LHC experiments and beyond. 

\section{The TimeSPOT sensor}
\label{sec:timespotsensors}
\begin{figure}[htbp!] 
\centering    
\includegraphics[width=0.5\textwidth]{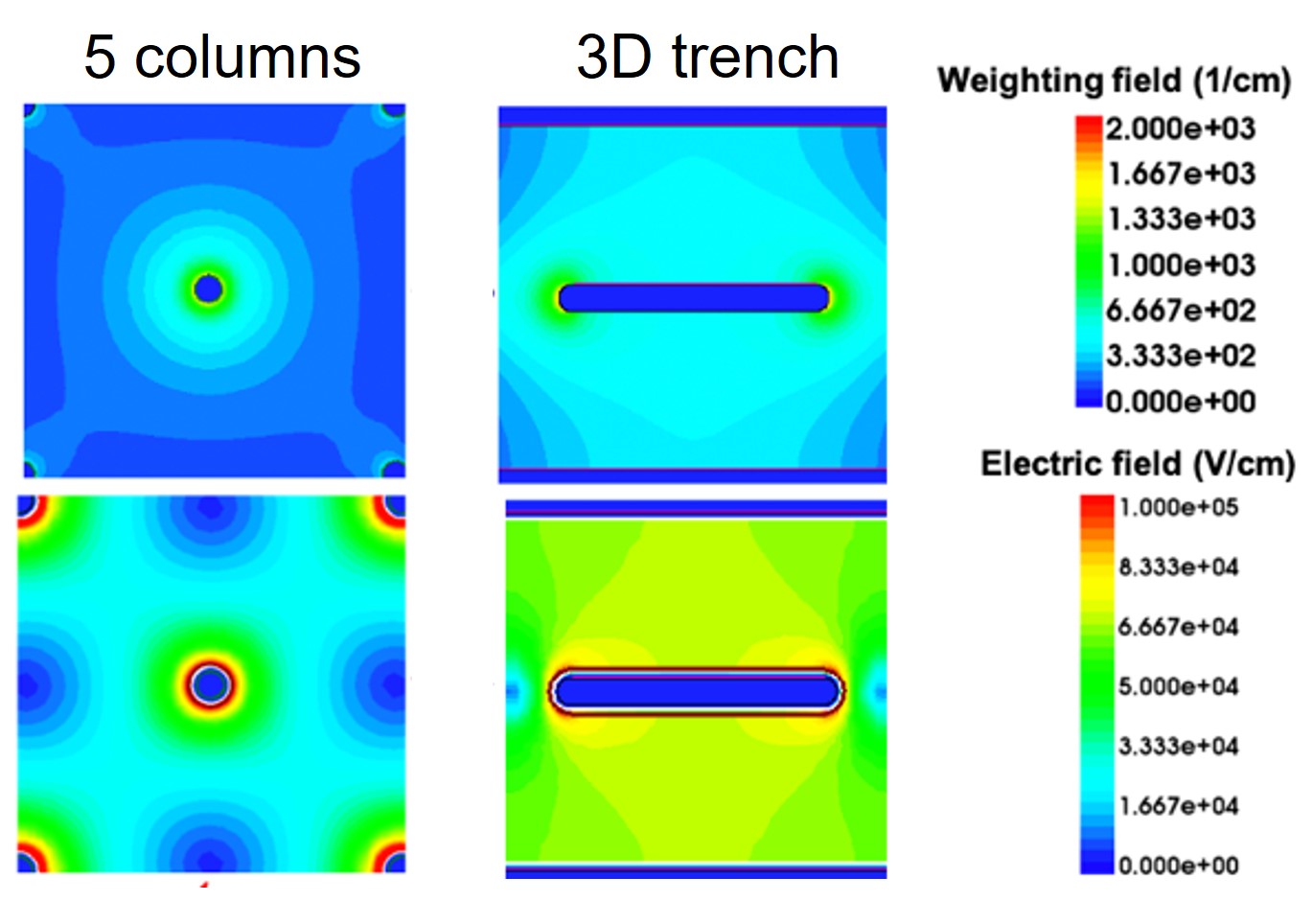}
\caption{TCAD simulations of weighting field and electric field for a 3D detector with a 5 columns electrodes geometry (left) and for the TimeSPOT 3D-trench geometry (right). -150 V bias voltage simulation.}
\label{fig:TCAD}
\end{figure}

\begin{figure}[htbp!] 
\centering    
\includegraphics[width=0.6\textwidth]{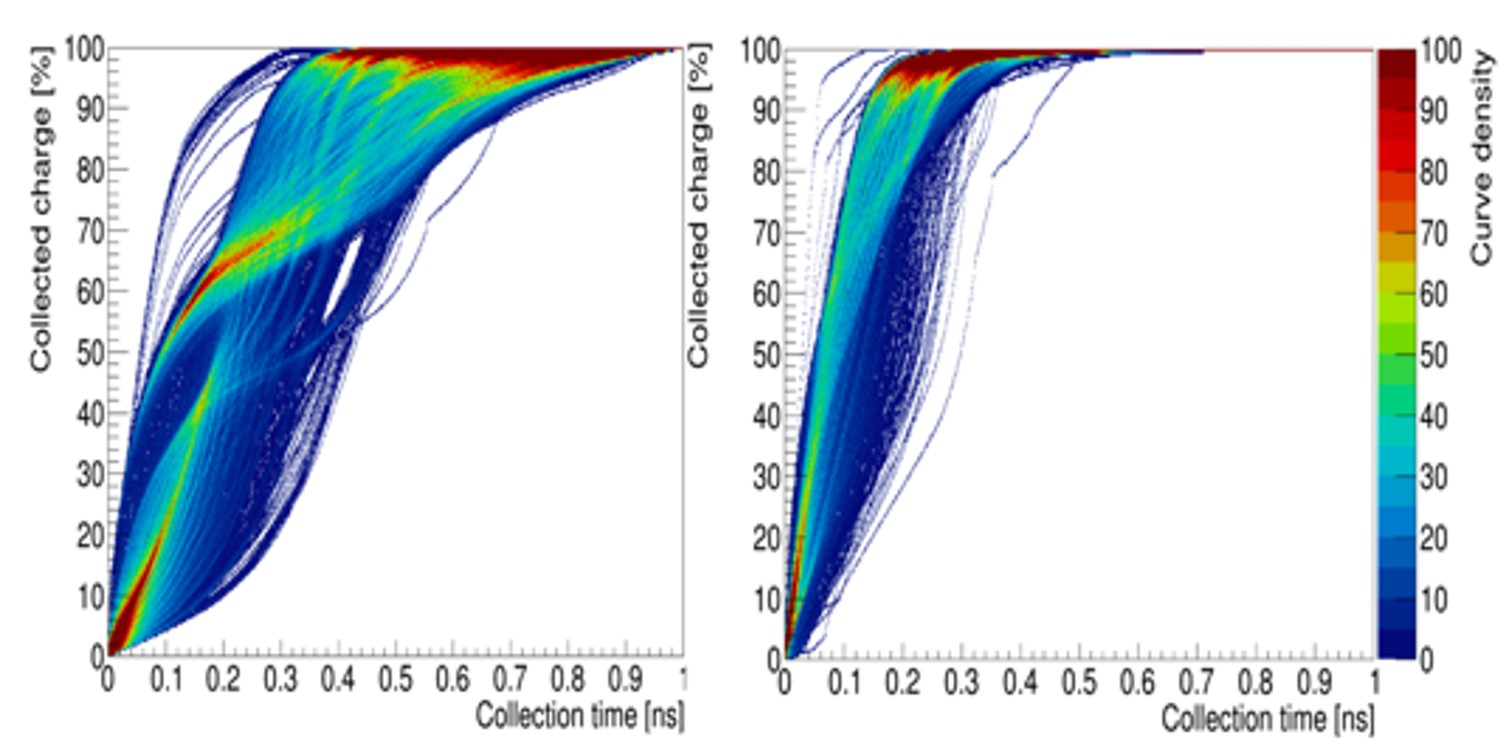}
\caption{Simulations of charge collection curves for MIPs uniform incidence for the five columns geometry (left) and for the trench geometry (right).}
\label{fig:TCODE}
\end{figure}

The TimeSPOT project, by means of a very complete simulation work, has designed new pixel structures based on the trench design~\cite{jinst}\cite{jinst2}. This configuration is based on trench electrodes that recreate a planar-like geometry. This electrode shape leads to a weighting field~\cite{Ramo}, uniform by design, and to a high and uniform electric field that give access to velocity saturation regime in almost the entire active area of the sensor~\cite{3dgeometry}. TCAD simulations allow to compare electric field and weighting field maps of a standard 5 columns 3D geometry and the trench geometry, that is shown in Figure~\ref{fig:TCAD}. The trench design, despite the other classic 3D geometry, presents more uniform electric and weighting field. As a result of this, Figure~\ref{fig:TCODE} shows that the simulated charge collection curves, obtained for Minimum Ionizing Particles (MIP) that uniformly crosses the pixels sensitive area, provide shorter and much more uniform charge collection time for the trench design.

\begin{figure}[htbp!] 
\centering    
\includegraphics[width=0.5\textwidth]{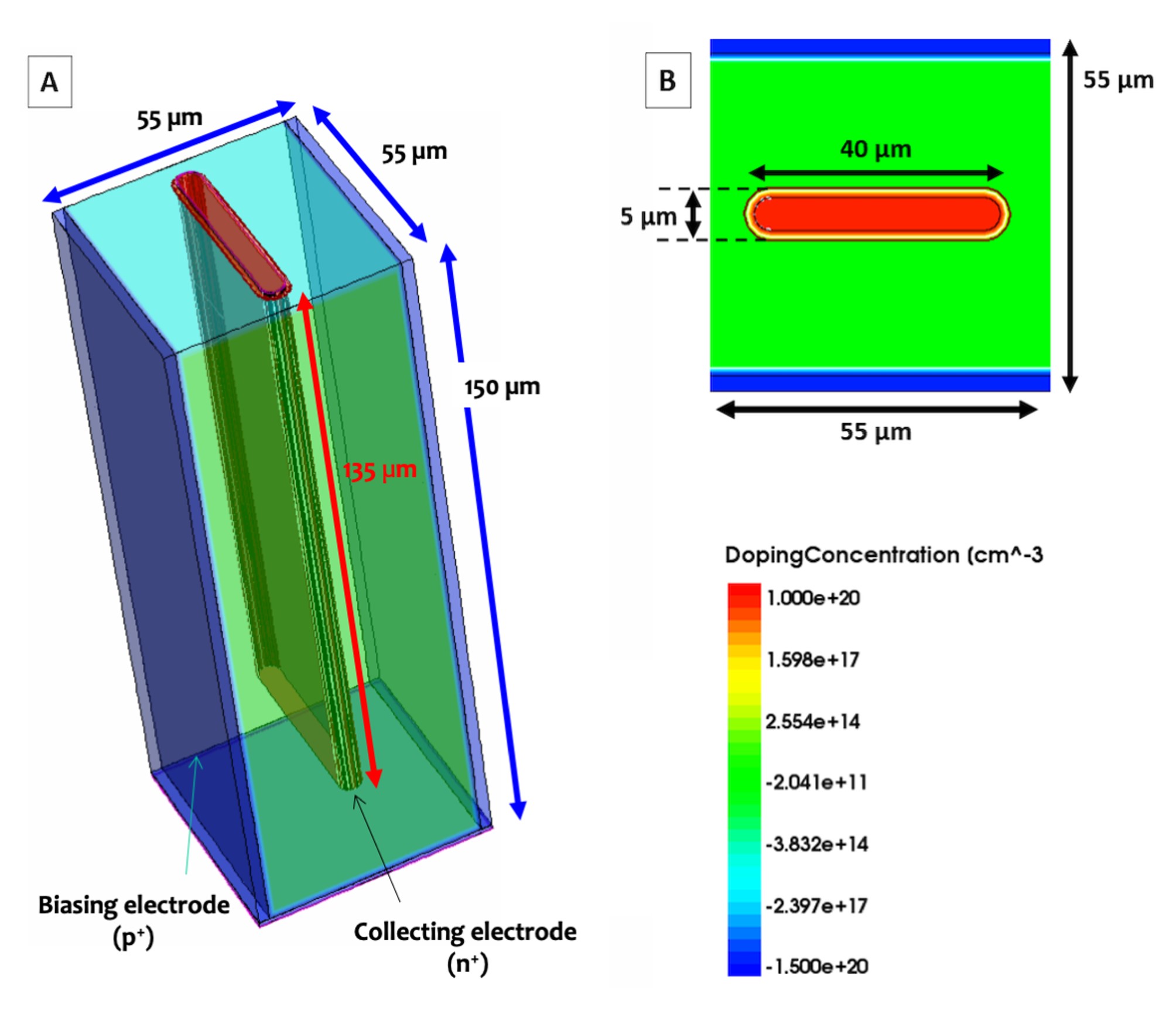}
\caption{The TimeSPOT 3D trench silicon pixel sensor.}
\label{fig:tpixel}
\end{figure}

The TimeSPOT pixel, shown in Figure~\ref{fig:tpixel}, has a dimension of $55 \times55\,\mu$m$^2$ with an active thickness of 150 $\mu$m. In each pixel a 40 $\mu$m long n++ trench, the readout electrode, is placed between two p++ trenches used for the bias. 
The readout electrode is only 135 $\mu$m deep and under the pixel there is a support wafer of 350 $\mu$m thickness.

Two batches of sensors were produced by FBK
(Fondazione Bruno Kessler, Italy) in 2019 and 2021 using the Deep Reactive Ion Etching Technique (DRIE) Bosch process. Several test structure were fabricated: single pixels, double pixels, pixel-strips

\section{Tested devices and front end amplifier board}
\label{DUT}
\begin{figure}[htbp!] 
\centering    
\includegraphics[width=0.6\textwidth]{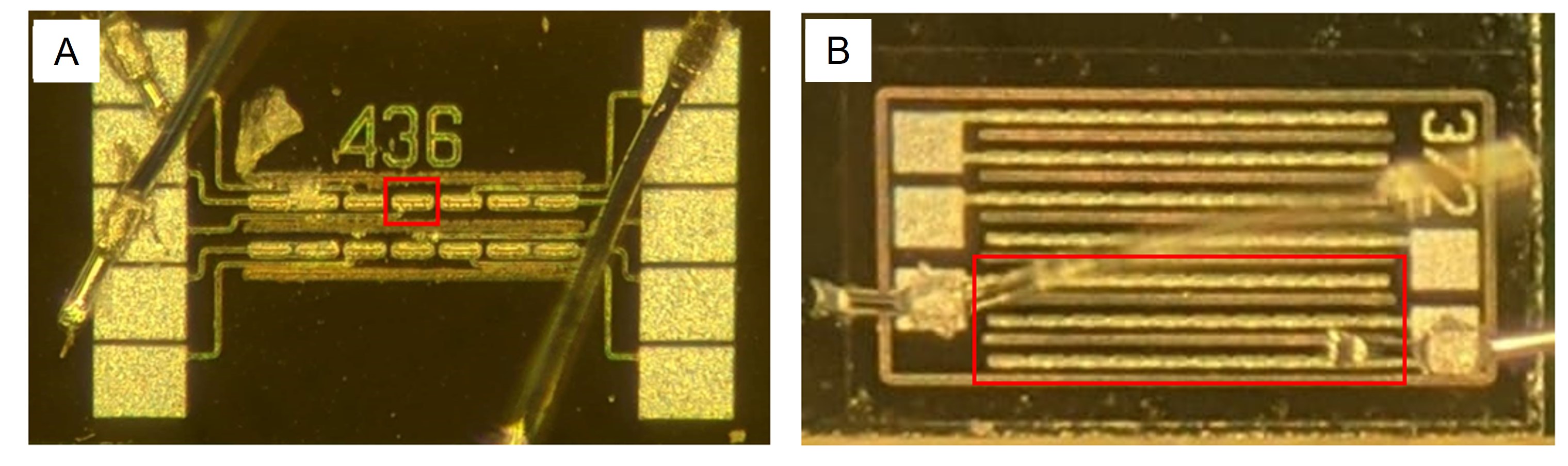}
\caption{Pictures of some of the 3D trench silicon pixel test structures used in this work. For each structure the active
area is outlined in red. (A) Single pixel sensor; (B) triple strip sensor: 30 pixels located in three adjacent rows. }
\label{fig:strutture}
\end{figure}

The sensors characterized in this paper, reported in Figure~\ref{fig:strutture}, are single pixels and triple-strips structures in which 30 pixels are read by the same amplifier channel. The tested devices are both non-irradiated and irradiated sensors up to a fluence of $2.5 \cdot 10^{16}\,1\,MeV\, n_{eq}\,cm^{-2}$.
To amplify the fast sensors currents a custom-made front-end electronics board is used. It is composed of two stages of transimpedance amplifier (TIA), made with low noise and high bandwidth silicon germanium BJTs and features a very low jitter, for MIP equivalent charge deposition, of about 6 ps~\cite{tiajitter}.

\section{Beam test setup}
\begin{figure}[htbp!] 
\centering    
\includegraphics[width=0.6\textwidth]{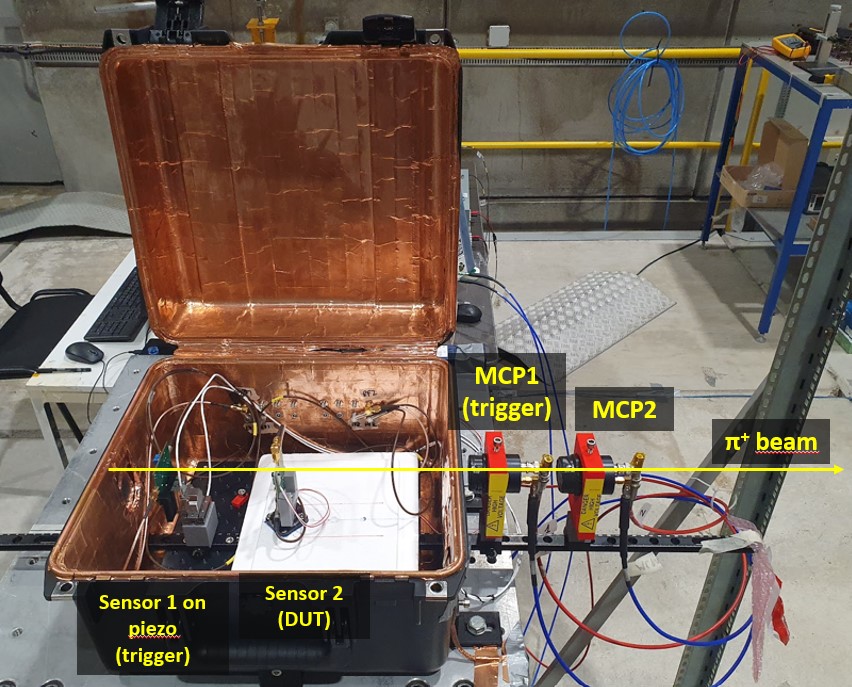}
\caption{The setup used for the measurements described in this work. The sensors mounted on their FEE boards inside the RF shielded and light tight box and the two MCP-PMTs downstream.}
\label{fig:setup}
\end{figure}

The sensors are tested in the H8 line of SPS with a 180 GeV/c $\pi^+$ beam. 
The experimental setup, shown in Figure~\ref{fig:setup}, consists of 2 MCP-PMTs aligned to the beam providing, when used together, a time resolution of about 4 ps. The Device Under Test (DUT) is placed at the center of a black RF shielded box, used to reduce electronics pick up on the DUT and also to operate the sensors in the dark.

The DUT mount allows to tilt the board in order to test sensors with various tilting angles with respect to the normal incidence. A trigger sensor is mounted upstream on two piezoelectric stages allowing to fine align two pixels. The alignment of two small 3D sensors allows to perform unbiased characterizations without using any amplitude selection on the DUT signals and also to perform efficiency measurements.
With this setup it is also possible to test irradiated sensors at low temperature, from -40° to -20° C placing a polystirene box, filled with dry ice, inside the black box. A temperature sensor in thermal contact with the amplifier board is used to check the DUT temperature during the measurements.

The sensors signals are recorded via $8$~GHz analog bandwidth, $20$~GSa/s 4 channels oscilloscope, the Rhode\&Schwartz RTP084. A typical event consists of four signals, two from the MCP-PMTs and two from the 3D sensors; the signals are then analyzed offline in order to measure their amplitude and Time of Arrival (ToA).

\section{Analysis and results}
In this section the analysis of the recorded signals and the results are presented.
The analysis consists of measuring the DUT signal amplitude, measured as the peak of the waveform subtracted by the baseline, calculated event by event, and the signal ToA that is evaluated with three different methods.
A leading edge algorithm gets the time in which the signal exceeds 15 mV, a linear interpolation is done around the threshold. The Spline method, a CFD based algorithm, gets the time at the 20\% of the signal amplitude; the signal is interpolated with splines in order to reduce the jitter due to the oscilloscope sampling frequency.
The last method is the Reference method, known in literature as the Amplitude Rise time Compensated discriminator (ARC)~\cite{knoll}: in this case a delayed copy of the signal is subtracted to itself, then the ToA is given as the time in which this new signal reaches the 50\% of its amplitude. The benefit of the Reference method is to reduce the time jitter due to rise time fluctuations.

\subsection{Amplitude}
\begin{figure}[htbp!] 
\centering    
\includegraphics[width=0.8\textwidth]{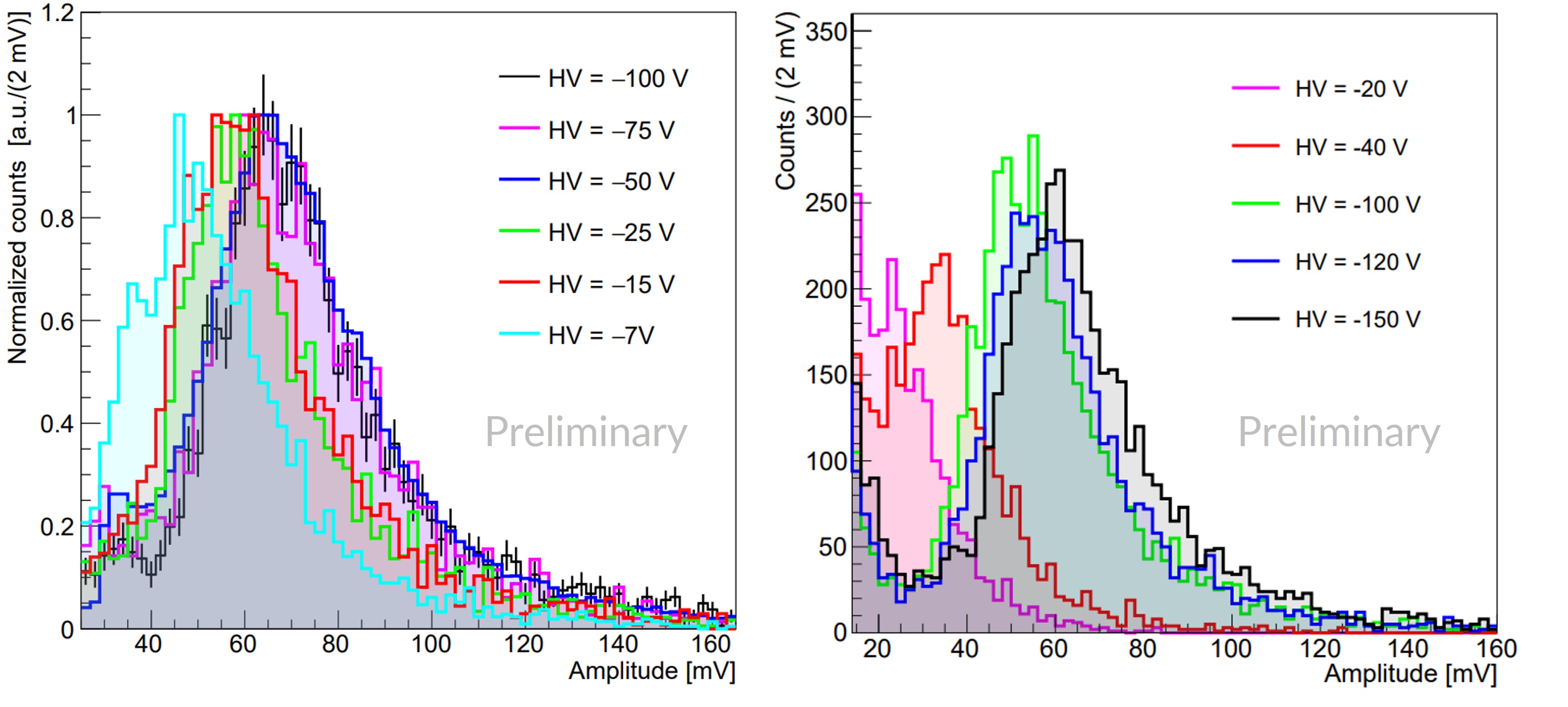}
\caption{Amplitude distributions for non-irradiated (left) and irradiated at a fluence of $2.5 \cdot 10^{16}\,1\,MeV\, N_{eq}\,cm^{-2}$ (right) 3D trench single pixels for different bias voltages.}
\label{fig:unirrampl}
\end{figure}

The amplitude measurements allow to evaluate and compare the overall charge collection performances of the non-irradiated and irradiated sensors. Figure~\ref{fig:unirrampl} shows the amplitude distributions for a non-irradiated 3D trench single pixel operated at different bias voltages for normal beam incidence. A good charge collection performance even at very low bias voltages (-7 V) is measured. The measurements of the irradiated pixel at the fluence of $2.5 \cdot 10^{16}\,1\,MeV\, n_{eq}\,cm^{-2}$ confirm the high radiation hardness of 3D devices, in fact increasing the bias voltage higher than -100 V the performances of the non-irradiated pixel are fully recovered.

\subsection{Time Resolution}
\begin{figure}[htbp!] 
\centering   
\includegraphics[width=0.9\textwidth]{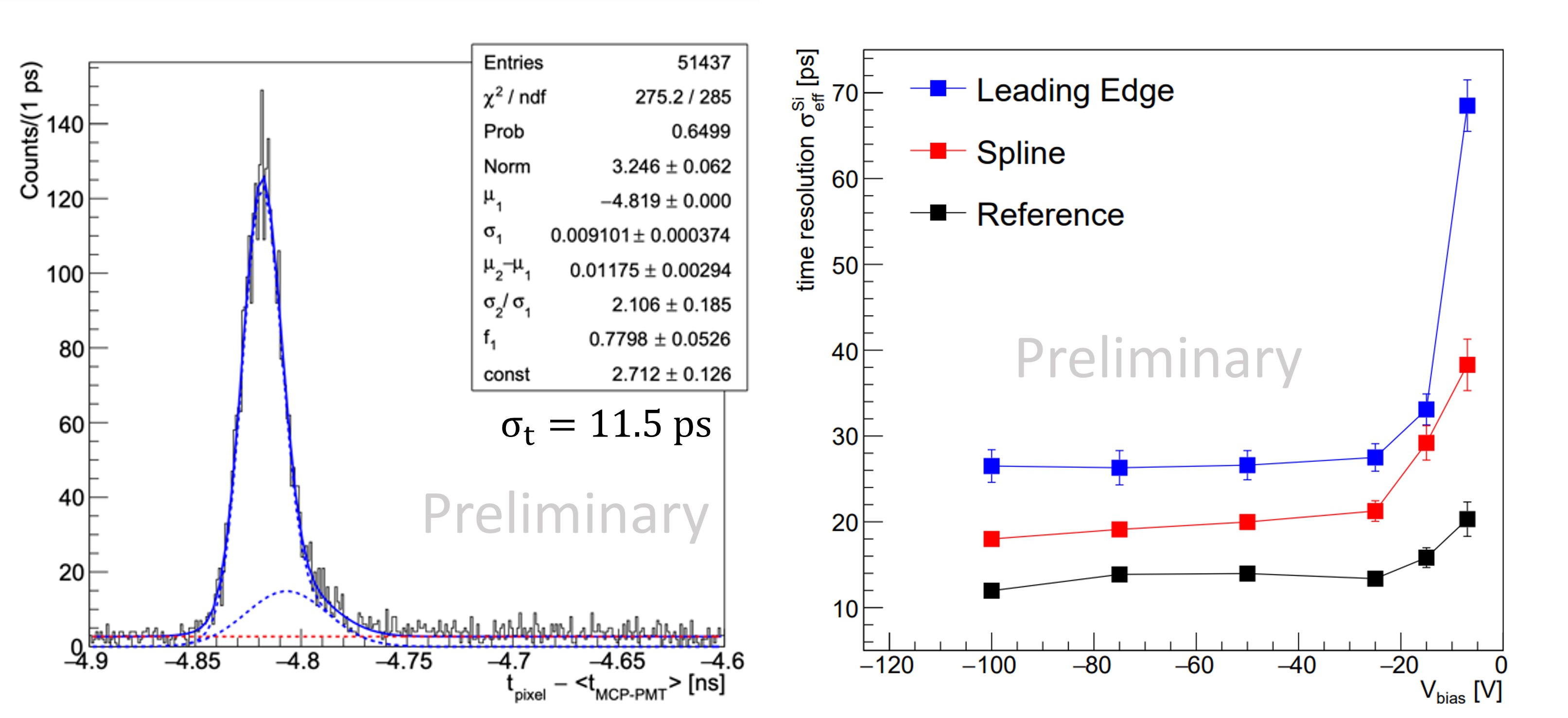}

\caption{(left) Time distribution of a non-irradiated 3D trench single pixel at -100 V. (right) Time resolution as a function of the applied bias voltage for the same pixel.}
\label{fig:timedistribution}
\end{figure}

The time resolution is estimated measuring multiple times the DUT ToA with respect to the absolute time measured with the time reference detector and measuring its fluctuation. 
Figure~\ref{fig:timedistribution} shows the distribution of a non-irradiated 3D trench single pixel with respect to the time of the MCP-PMTs. To estimate the time resolution a two gaussians fit is performed on the distribution.

The two gaussians fit function allows to take into account the small asymmetric tale of late signals. Then the time resolution is quoted as the standard deviation of the distribution, calculated as

\begin{equation}
\sigma_t^2 = f_1(\sigma_1^2+\mu_1^2)+(1-f_1)\cdot(\sigma_2^2+\mu_2^2)-\mu^2,
\end{equation}

where $f_1$ is the fraction of the Gaussian core and $\mu$ is

\begin{equation}
\mu = f_1\mu_1+(1-f_1)\cdot\mu_2.
\end{equation}

\begin{figure}[htbp!] 
\centering    
\includegraphics[width=0.44\textwidth]{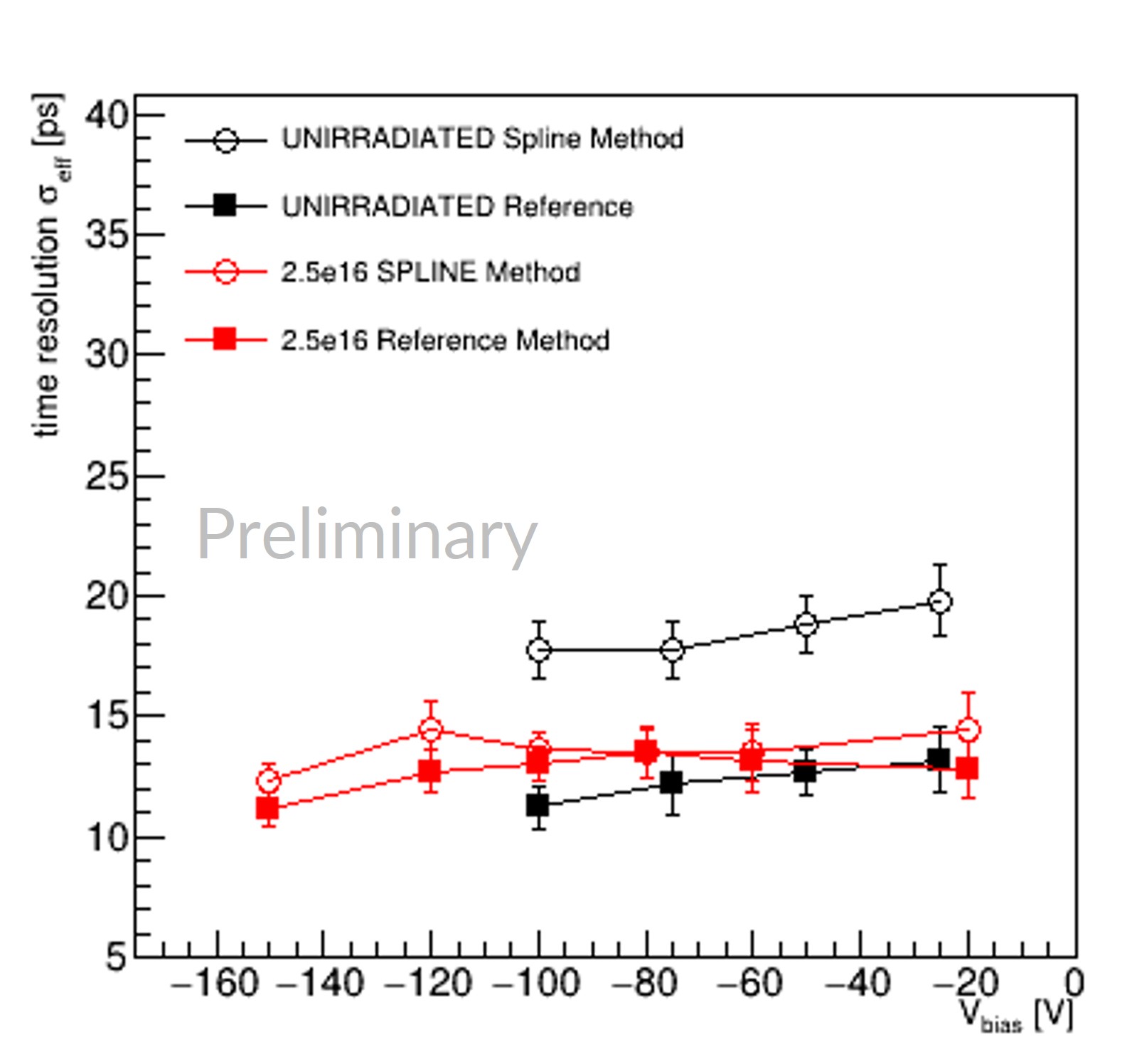}
\includegraphics[width=0.54\textwidth]{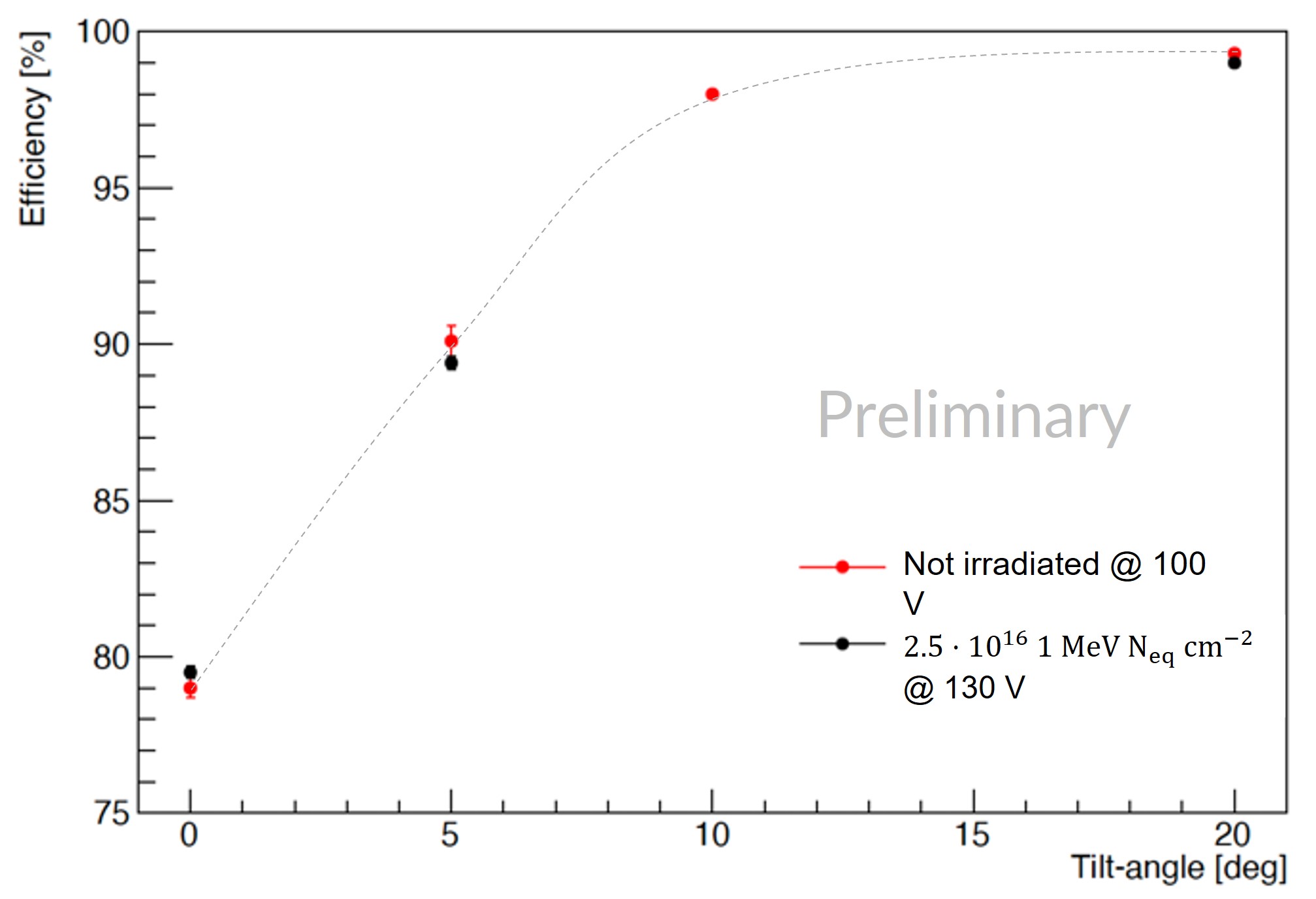}
\caption{(Left) Single pixel time resolution as a function of the bias voltage for (red) irradiated sensor and (black) non-irradiated.
(Right) Triple strip detection efficiency for a non-irradiated  (red) and irradiated (black)  as a function of the tilt angle with respect to the normal incidence.}
\label{fig:irrtime}
\end{figure}

This leads to a time resolution of 11 ps at a bias voltage of -100 V using the Reference method.
Figure~\ref{fig:timedistribution} (right) shows no evidence of a strong dependence of the time resolution to the bias voltage down to -20 V for all the proposed methods.
Excellent timing performances, below 30 ps, are obtained even with the leading edge algorithm, a simple discriminator implementable in ASICs, with low area impact.
The trench pixel irradiated at $2.5 \cdot 10^{16}\,1\,MeV\, n_{eq}\,cm^{-2}$ shows very similar performances to the non-irradiated pixel featuring a time resolution of 10 ps at a bias voltage of -150 V. Also in this case, no significant time resolution degradation emerges for biases up to -20 V, Figure~\ref{fig:irrtime} (left).

\subsection{Efficiency}
The trenches electrodes of the TimeSPOT pixels are not active volume so if a particle channels inside a trench it will not be detected, decreasing the pixel detection efficiency. A typical solution to recover 3D sensor geometrical efficiency is to operate them with a tilted angle with respect to the normal incidence. Dedicated measurements of 3D trench pixel detection efficiency are presented in this paper. 
In order to measure the detection efficiency a 3D trench single pixel is aligned to the DUT, that in this case is a 3D trench triple-strip structure. This configuration allows to estimate the efficiency as $\eta = N_{ts}/N_{tr}$, where  $N_{ts}$ is the number of events detected by the triple-strip sensor and $N_{tr}$ is the number of trigger (number of particles that geometrically cross the triple strip sensor).
The results are shown in Figure~\ref{fig:irrtime} (right), for both the non-irradiated and the irradiated at $2.5 \cdot 10^{16}\,1\,MeV\, n_{eq}\,cm^{-2}$ triple strips sensors. The efficiency increases as the tilting angle increases reaching 99\% detection efficiency for tilt angles higher than $10^\circ$. Also the irradiated sensor recovers the full efficiency, in this case measured at a bias voltage of -130 V.

\section{Conclusion}
New particle beam characterizations of 3D trench silicon pixel sensors are shown in this paper. 3D-trench pixel sensors have shown a 10 ps time resolution for MIP detection with a 99\% detection efficiency when operated at tilt angles higher than $10^\circ$ with respect to normal incidence. The same results are obtained for 3D trench sensors irradiated at $2.5 \cdot 10^{16}\,1\,MeV\, N_{eq}\,cm^{-2}$, indicating that the radiation hardness limit of this technology it is not reached yet at this fluence. 
The results presented in this paper are the first beam test timing characterization of highly irradiated 3D silicon pixel sensors and they prove that 3D trench sensors are one of the best candidate for 4D tracking detectors at very high fluences in high-energy-physics colliders.

\acknowledgments

This work was supported by the Fifth Scientific Commission (CSN5) of the Italian National Institute
for Nuclear Physics (INFN), within the Project TIMESPOT and by the ATTRACT-EU initiative,
INSTANT project. The authors wish to thank Heinrich Schindler that, as LHCb beam test coordinator, has given logistic support for the setup installation at the SPS H8 line.



\begin{thebibliography}{99}

\bibitem{ftdr}
LHCb Collaboration, \emph{Framework TDR for the LHCb Upgrade II}, CERN-LHCC-2021-012, LHCB-TDR-023, 2021.

\bibitem{jinst}
L. Anderlini et al., \emph{Intrinsic time resolution of 3D-trench silicon pixels for charged particle detection}, JINST 15 P09029, 2020.

\bibitem{jinst2}
D. Brundu et al., \emph{Accurate modelling of 3D-trench silicon sensor with enhanced timing performance and comparison with test beam measurements}, JINST 16 P09028, 2021.

\bibitem{radhard}
M. Manna et al., \emph{First characterisation of 3D pixel detectors irradiated at extreme fluences}, Nucl. Instrum. Meth. A979 (2020) 164458.

\bibitem{Ramo}
S. Ramo, \emph{Currents Induced by Electron Motion}, in Proceedings of the IRE, vol. 27, no. 9, pp. 584-585, Sept. 1939.


\bibitem{3dgeometry}
A. Loi et al., \emph{Timing optimisation and analysis in the design of 3D
silicon sensors: the TCoDe simulator}, JINST 16 P02011, 2021.

\bibitem{tiajitter}
M. Aresti et al., \emph{Laboratory Characterization of Innovative 3D Trench-design Silicon Pixel Sensors Using a Sub-Picosecond Precision Laser-Based Testing Equipment}, 2020 IEEE Nuclear Science Symposium and Medical Imaging Conference (NSS/MIC), 2020.

\bibitem{knoll}
Glenn F. Knoll, \emph{Radiation Detection and Measurement}, Wiley, 2010.





\end{thebibliography}
\end{document}